# Generation and Tunability of Supermodes in Tamm Plasmon Topological Superlattices

*Tong Qiao [1], Mengying Hu [1], Xi Jiang [1], Qiang Wang [1], Shining Zhu [1] and Hui Liu [1,*]*

[1] National Laboratory of Solid State Microstructures, School of Physics, Collaborative Innovation Center of Advanced Microstructures, Nanjing University, Nanjing, 210093, People's Republic of China.

*E-mail: liuhui@nju.edu.cn

In this study, we propose and experimentally demonstrate a novel kind of Tamm plasmon topological superlattice (TTS) by engineering Tamm photonic crystals (TPCs) belonging to a different class of topology. Utilizing specifically designed double-layer metasurfaces etching on planar multilayered photonic structures, the TPC that supports the Tamm plasmon photonic bandgap is realized in the visible regime. Through the coupling of topological interface states existing between different TPCs, hybrid topological interface states of Tamm plasmon, called supermodes, are obtained that can be fully described by a tight-binding model. Meanwhile, we can achieve a tunable bandwidth of supermodes via varying the etching depth difference between double-layer metasurfaces. We show that the bandwidth decreases with the increase of etching depth difference, resulting in a nearly flat dispersion of supermodes with strong localization regardless of excitation angles. All the results are experimentally verified by measuring angular-resolved reflectance spectra. The TTS and supermodes proposed here open a new pathway for the manipulation of Tamm plasmons, based on which various promising applications such as integrated photonic devices, optical sensing, and enhancing light-matter interactions can be realized.

For many years, plasmonics has been acting as a hot research area in optics that offers a new route to manipulate light in the nanoscale.[1,2] For example, surface plasmon polaritons (SPPs), characterized by strong optical field localization and enhancement, are extensively exploited in the realms of integrated optics, nonlinear optics, optical sensing, molecular detection, and quantum optics.[2-4] However, the radiative or ohmic loss, causing the energy of SPPs dissipated and decaying quickly, hinders its practical application range.[5,6] In recent years, Tamm plasmon polaritons (TPPs),[7,8] which are optical states mainly located at the interface between a dielectric Bragg reflector (DBR) and a thin metal film, have aroused people's attention. Compared with ordinary SPPs, TPPs inherit the advantages of plasmons, but possess much less loss. More specifically, TPPs have both TE/TM modes simultaneously. The dispersions of TPPs are above the light cone so that it can be excited by free space light directly without the aid of diffraction gratings or prisms.[7-19] Therefore, field enhancement can be realized flexibly in all kinds of nanophotonic devices employing TPPs, featured by high local density and, thus, dramatically enhancing the light-matter interaction. These unique characteristics lead to various applications, such as perfect absorption,[9,10] Tamm plasmon lasers,[11-13] optical detection[14,15] and thermal emission,[16-18] enhancement of nonlinear effects,[19,20] and so on. In addition to the traditional use of metal films to achieve TPPs, researchers also utilized metasurfaces,[21,22] liquid crystals,[23] and microcavity exciton[24,25] for active tuning of TPPs. At the same time, the structures supporting propagating Tamm-like Bloch modes have been obtained via periodically patterning the metal layer on the top of dielectric multilayers.[26-28]

On the other hand, recently, topological photonics[29-31] has drawn a great deal of interest owing to its exotic properties. Especially topologically protected states, which are robust modes immune to optical scattering, have been intensively studied for the attainability of stable unidirectional transmission devices,[32] topological lasers,[33-35] and integrated and quantum photonic devices.[36,37] Topological states between distinct topological phases are explored in various systems, including plasmonic systems[38-48]. Introduction of the concept of topology to plasmonic systems not only provides a new route to the manipulation of surface wave propagation, but also simplifies the structures compared with those as previously reported.[40-42] Many intriguing phenomena have been realized in plasmonic systems, such as photonic Weyl points and Fermi arcs in magnetized plasma,[38-40] topological edge states in designer surface plasmon structures,[41-43] anomalous wave propagation of SPPs in topological metasurfaces,[44] and so on. In addition, engineering the chain of plasmonic nanoparticles such as metallic nanodisk arrays has also been proposed,[45-48] supporting topological edge states for the guiding and confinement of light. Especially, the work by *Chen*'s et al.[49] has reported the first observation of the Shockley-like surface states in an optically induced semi-infinite photonic superlattice. The Shockley-like surface states can be transformed into Tamm-like surface states by surface index variation. However, so far, very few studies have touched on the topological states in the TPPs context.[15] There is no report on the realization of engineering the chain or the lattice of TPPs, which can be a versatile integration platform with high tunability and beneficial for exploiting novel topological phases in plasmonics.

In this paper, we establish a Tamm topological superlattice (TTS) composed of alternative Tamm photonic crystals (TPCs) with distinct topological properties in the visible regime. Through meticulous design, we acquire controllable couplings between Tamm topological interface states, which are particularly confined modes mainly located at the boundary of adjacent TPCs and serves as an indispensable element to construct TTS supporting collective modes, called as supermodes[50-52]. The dispersion of supermodes resides in the Tamm photonic bandgap, which can be fully investigated by the tight-binding approach. By changing the geometric parameters of TTS, a variable coupling strength can be adopted to realize a varying supermodes dispersion with tunable bandwidths. Particularly, the dispersion associated with supermodes is nearly flat under some specific conditions, indicating a high density of optical states, which are desirable and useful for applications such as Tamm plasmon lasers,[11-13] enhanced spontaneous radiation[13,53] or plasmonic sensors.[14,15] All the band structures are experimentally mapped by measuring angle-resolved reflection spectra and agree well with calculations.

**Results and Discussion**

Figure 1a shows the schematic of TPC which is composed of a PC and double-layer metal gratings.[54] The designed PC structure comprises 18 unit cells, each of which consists of three layers labeled as A/2-B-A/2. Here, layer A is $HfO_2$ and layer B is $SiO_2$. The thickness of $HfO_2$ layer is 84 nm and that of the $SiO_2$ layer is 115 nm with refractive indices of 2 and 1.46, respectively.[55] As shown by red dashed lines in Figure 1a, the period of PC's unit cell is $\Lambda = 199$ nm. Instead of utilizing a thin metallic film as in previous Tamm devices, we pattern double-layer metal gratings on the top of PC. The grating is etched from a silver film with a period $L$ and strip with $w$. The etching depth in Figure 1a is defined as $h$. The double-layer metal gratings (yellow) consist of two parts: upper gratings labeled as "$a$" and lower gratings labeled as "$b$". The unit cell composes of three parts: "$b/2$", "$a$" and "$b/2$". Compared with the previously reported TPCs in the near-infrared region[28] with inevitable energy leaking from the slits of gratings, our scheme utilizing double-layer metal gratings can prevent the energy of the Tamm confined mode from leaking into the air, which may cause a large radiative loss, and thus the TPCs working in the visible regime are experimentally available (see details in Supporting Information, Figure S6).

In the experiment, we fabricate the periodical structures with a period $L$ = 400 nm and the strip width along the periodical direction is $w$ = 160 nm (detailed process of synthesis of TPCs referred to Methods). The etching depth $h$ is 85 nm. The scanning electron microscope (SEM) top view of the TPC sample is shown in Figure 1b. To experimentally demonstrate band structures of the TPC, we introduce the angle-resolved microspectroscopy system[56] to measure reflection spectra of the sample, as illustrated in Figure 1c (details seen in Methods). The information on the radiation field of the sample in the Fourier space or momentum space is the same as that on the back focal plane of the objective lens. Due to the fact that the spectrometer is conjugated with the back focal plane of the objective lens, we are capable of getting get the Fourier-transformed spectra. Then we can map band structures of the sample in the entire First Brillouin zone within the visible regime by altering the position of the scanning fiber on the incident plane and angle-resolving plane, which are both conjugated to the back focal plane of the objective lens.

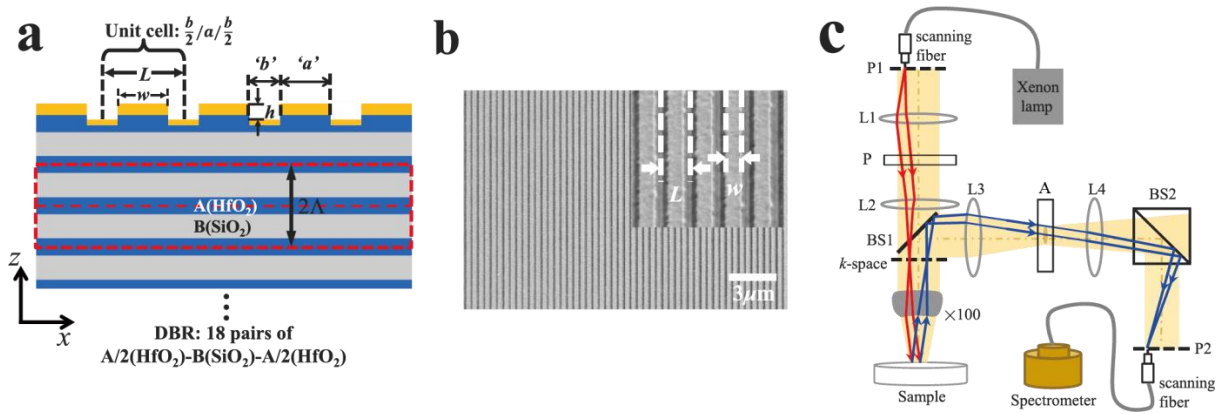

**Figure 1.** Schematic of TPC and experimental setup. (a) Schematic structure for TPC. The double-layer Ag gratings (yellow) consist of two parts: upper gratings labeled as "$a$" and lower gratings labeled as "$b$". The unit cell composes of three parts: "$b/2$", "$a$" and "$b/2$". The bottom PC is made up of alternating layers of two dielectric materials, denoted as A($HfO_2$) and B($SiO_2$). Two periods of PC are marked by red dashed lines. (b) SEM top view of the TPC sample with period $L$ and strip width $w$, marked by white dashed lines in the inset picture. (c) Setup of the angle-resolved reflection spectrum system. Red (blue) rays correspond to the incident (reflected) light. BS, beam splitter; L, lens; P, polarizer; A, aperture; P1, incident plane; P2, angle resolving plane.

In order to get dispersion of the TPC above, we numerically calculate the TE-polarized (y-polarized in Figure 1a band structures of the TPC by COMSOL Multiphysics (details referred to Methods). The calculated results are given as black circles in **Figure 2a**, in which the gray region refers to projected bulk bands of the PC. The white region corresponds to the bandgap of PC and all band structures shown in Figure 2a are above the light line. Due to backfolding of the dispersion of Tamm modes induced by the periodical structures, a photonic bandgap of TPC is formed at the edge of the First Brillouin zone, which is contained inside the bandgap of PC. Note that the artificial periodic condition also accounts for the fold of PC's bulk bands, in which a part of the folding bands covers the frequency range of PC's bandgap and even overlaps with the dispersion of Tamm modes. Consequently, the dispersion of Tamm modes will have some ripples attributed to the inevitable interaction between the Tamm modes and folding bands of PC.

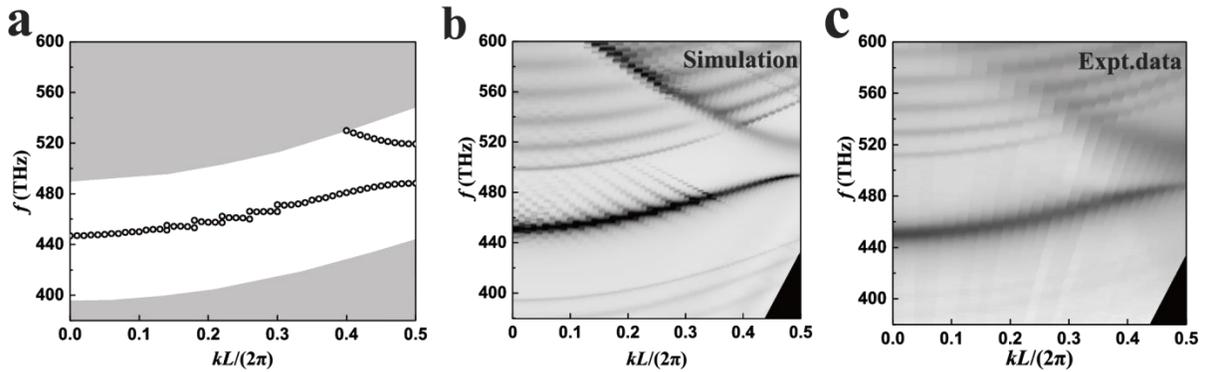

**Figure 2.** Dispersion of TPC. (a) Band structures of TPC for TE polarization. Gray Region indicates the projected bands of the PC. Black circles calculated by COMSOL represented TPC modes inside the PC's bandgap, which result in a photonic bandgap of TPC. (b) TE-polarized dispersion of TPC calculated by angle-resolved reflection spectra. Tamm modes are highlighted by the reflection dips inside the bandgap of PC. (c) Experimentally measured dispersion by angle-resolved reflection spectra in Figure 1c.

Moreover, the designed dispersion of Tamm modes and TPC's bandgap in Figure 2a are located above the light line, which means that they are readily available by the far-field measurement, such as angle-resolved reflection spectra. We achieve the dispersion of the TPC by virtue of Finite-Difference Time-Domain (FDTD) simulations (details referred to Methods). The simulated result of TE-polarized band structures for the TPC is given in Figure 2b, in which dips in the gap verify the existence of Tamm modes, matching well with the calculated curve of TPC dispersion in Figure 2a. Experimentally, we measure angle-resolved reflection spectra harnessing the system sketched in Figure 1c and the results of band structures under TE polarization are shown in Figure 2c. A TPC's bandgap with a width of 27.4 THz emerges, equivalent to 25% of the PC's bandgap, showing great conformance with simulation. Furthermore, the experimental adjustability of the etching depth $h$ endows us with the ability to tailor the width of the TPC bandgap facilely. In particular, it can be seen that the larger the etching depth $h$ is, the more confined the Tamm mode becomes, resulting in a larger TPC's bandgap (detailed results of a varying TPC's bandgap with different $h$ can be seen in Supporting Information, Figures S2 and S3).

Based on the TPC above, we start to construct a TTS from forming a Tamm topological interface state. Two kinds of TPCs, whose unit cell is labeled as "p": (a/2-b-a/2) and "q": (b/2-a-b/2), respectively, are shown in the upper panel of **Figure 3a**. Here "a" and "b" are the upper and lower gratings, as marked in Figure 1a with the same parameters. We use $p_n$ ($q_n$) to represent the TPC, including n unit cells of p (q). Both TPCs $p_n$ and $q_n$ are binary centrosymmetric and have identical compositions. They have the same band structures, but discriminative band topologies, owing to the different unit cell origins.[57,58] Therefore, when we combine them together, a boundary forms and there will be a Tamm topological interface state within the bandgap (topological origin of the Tamm interface state can be referred to Supporting Information, Figure S1).[57-59] As shown in the middle panel of Figure 3a, electric profile of the interface state is confined at the interface between two TPCs, in both horizontal and perpendicular directions. Assisted by COMSOL, the eigenfrequency of a single interface state, indicated by $\hbar\omega$ can be obtained as $\hbar\omega = 2.07$ eV, which is inside the common bandgap of two TPCs.

Next, we investigate the coupling effect between two adjacent Tamm interface states, which is indispensable for constructing the TTS subsequently. If we put three TPCs together, for example, two TPCs $q_4$ are separated by a TPC $p_4$, we will get two interfaces, either of which can support a topological interface mode. These two interface modes can couple with each other and form two hybrid modes subject to a frequency splitting: one mode is a symmetric (S) and the other is antisymmetric (AS). Here, the symmetric types are defined by the symmetry of the electric field which uses the center of $p_4$ as the reference point. The electric field profiles obtained from simulations are shown in the lower panel of Figure 3a. The boundaries of adjacent TPCs are marked by black dashed lines and the dotted line indicates the reference point. We can derive the associated coupling coefficient denoted by $t$ between two interface modes from the frequency splitting $t = (\omega_S - \omega_{AS})(2\omega_0)^{-1}$, where $\omega_S$ ($\omega_{AS}$) is the frequency of the symmetric (antisymmetric) mode and $\omega_0$ corresponds to the single mode. Here, we numerically calculate the coupling coefficients for two cases as TPC $p_4$ and TPC $q_4$ serve as coupling channels, respectively, by setting up the configuration that TPC $q_4$ ($p_4$) sandwiched by two TPCs $p_4$ ($q_4$) for the former (latter). The coupling coefficient via TPC q4 is $t_{q4} = -0.0041$ while that via p4 is $t_{p4} = 0.0054$. The reason why the signs of $t_{q4}$ and $t_{p4}$ are opposite is the accumulative phases for the coupling channels p4 and q4 are distinct, which is consistent with previous works.[58-60]

According to the analysis above, the TTS is built up of repeated TPCs p4 and q4, in which consecutive topological interface states are attainable and form a TTS shown in the upper panel of Figure 3b. Each Tamm interface state can be regarded as an artificial atom that hybridizes with its nearest neighbor, and the TTS is analogous to a diatomic chain shown in the lower panel of Figure 3b, which can be described by a tight-binding model. The diatomic chain is formed by alternatively connecting red and blue atoms characterized by the same eigenfrequency $w_1 = w_2$, which is intrinsically protected by the chiral symmetry of our system. A red (blue) atom represents the interface state existed at the interface of p4 (q4) and q4 (p4). The bonding between atoms corresponds to the coupling channels p4 or q4. Such a diatomic chain can be described by the SSH Hamiltonian:

$$H = \hbar\omega_1 \sum_i a_i^+ a_i + \hbar\omega_2 \sum_i b_i^+ b_i + \sum_i (t_1 a_i^+ b_i + t_2 a_{i+1}^+ b_i) + h.c. \quad (1)$$

Here $t_1$ and $t_2$ are the intradimer and interdimer coupling coefficients, respectively. $a_i^+$ and $b_i^+$ denote photonic creation operators of interface states inside the i-*th* unit cell. Thus, the dispersion of collective modes akin to that of a diatomic chain, called supermodes, can be obtained from the diagonalized Hamiltonian in eq 1, which is given as a red curve in Figure 3d. The dispersion of the Tamm modes, which corresponds to the common band structure of TPCs p and q is also depicted by black circles in Figure 3d. The supermodes are located inside the bandgap of TPC. To further testify our argument above, we calculate the supermodes dispersion using COMSOL, which is represented by blue circles and in accordance with the tight-binding result. Notice that since the length of unit dimer is 8*L*, the corresponding First Brillouin region of the TTS becomes one-eighth of that of the TPC in Figure 1a. Thus we exploit the dispersion of supermodes to the extended Brillouin region, as shown in Figure 3d, where vertical black dashed lines mark edges of extended Brillouin zones. Part of supermodes' dispersion outside the red dashed rectangle in Figure 3d covers the frequency range of projected bands of PC under small wave vectors ($kL/(2\pi) < 0.3$), which are represented by light blue circles (red line) and cannot be detected in the angle-resolved reflection spectra experiment because of mode leaking.

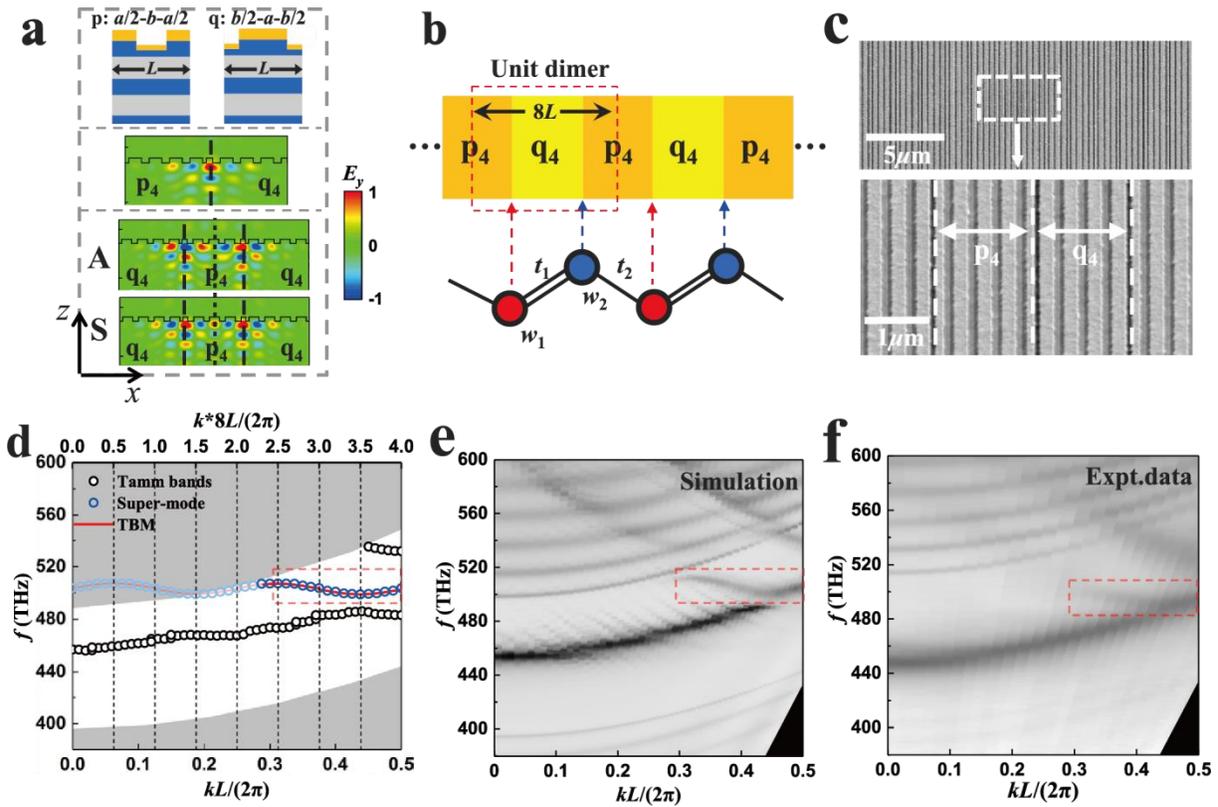

**Figure 3**. Realization of a Tamm topological superlattice (TTS). (a) The upper panel: two kinds of TPCs labeled as p ($a$/2-$b$-$a$/2) and q ($b$/2-$a$-$b$/2). The middle panel: electric profiles of a single Tamm interface mode. The lower panel: electric profile of the symmetric (S) and antisymmetric (AS) hybrid modes. Black dashed lines mark the boundaries of adjacent TPCs and the dotted line indicates the reference point. (b) The TTS is shown in the upper panel, and the schematic representation of effective dimerized model is given in the lower panel. The length of unit dimer composed of $p_4$ and $q_4$ is $8L$. (c) SEM top view of the TTS sample. The enlarged picture in the lower panel shows the supercell comprised $p_4$ and $q_4$. (d) Band structures of TTS for TE

polarization. Blue circles (red line) are (is) employed to display the numerical (analytical) result corresponding to supermodes given by COMSOL (Tight-Binding Model, TBM). Part of supermodes' dispersion covered by the projected bands of PC is represented by light blue circles (red line) outside the red dashed rectangle. Black dashed lines mark the boundaries of the extended Brillouin zones for supermodes. (e, f) Simulated (e) and measured (f) dispersion of TTS by angle-resolved reflection spectra under TE excitation. The red dashed rectangles in (d)-(f) mark the locations of supermodes inside the TPC's bandgap.

In experiments, we utilize the angle-resolved reflection spectrum to investigate the dispersion of TTS. The TTS sample is fabricated with the same parameters as the TPC above. Figure 3c shows SEM top view of the TTS sample (upper panel) and the unit cell of TTS comprised of $p_4$ and $q_4$, named as the supercell (magnified in lower panel). The simulated and experimental results of reflection spectra in terms of incident light with TE polarization are given in Figures 3e and f, respectively. Within the TPC's bandgap, we observe a dispersion curve of supermodes unambiguously inside the red dashed rectangles, as predicted by the tight-binding model (TBM). Apart from the unit dimer of TTS, which is made up of $p_4$ and $q_4$ ($p_4$-$q_4$), simulated dispersion of TTS with other unit dimer configurations, such as $p_2$-$q_6$ or $p_6$-$q_2$, can be seen in Supporting Information (Supporting Information, Figure S5). Moreover, it should be noted that supermodes are a novel variety of artificial collective modes and they reveal topological features of the TTS, which differ from edge states in zigzag or other tight-binding lattices. This ingeniously engineered TTS offers a new platform for the manipulation of TPPs and hence can be applied to achieve local field enhancement.

More intriguingly, the supermodes dispersion can be flexibly manipulated by tuning the etching depth $h$ of period structures fabricated on the PC. In practice, we fabricate a series of TTS samples featured by $h$ = 60 nm, 70 nm, 85 nm and 100 nm. The experimental band structures of supermodes are shown in **Figure 4a** (detailed simulated and experimental band structures of TTS under variable etching depth are given in Supporting Information, Figure S4). The bandwidth of supermodes dispersion varying as a function of $h$ is retrieved and plotted in Figure 4b, which can be reduced from 15.1 to 5.74 THz when $h$ increases from 60 to 100 nm.

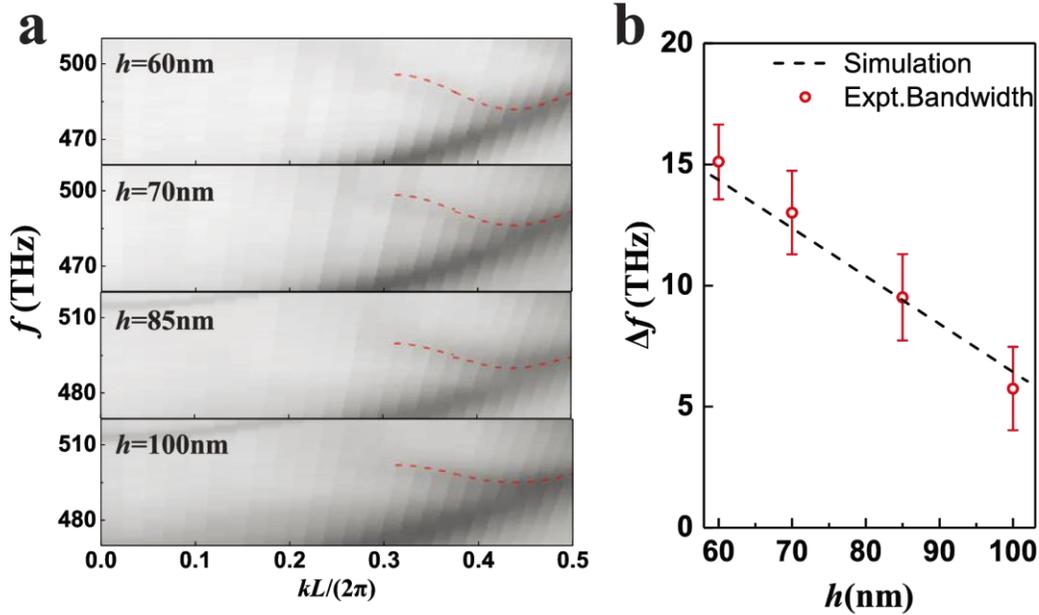

**Figure 4**. Dispersion of supermode and its tunability. (a) Part of the experimental band structures of the supermode with variable etching depths $h$ = 60, 70, 85 and 100 nm, respectively. Red dashed lines are dispersions calculated by tight-binding model. (b) Simulated (black dashed line) and experimental (red) bandwidth of the supermodes attained by different etching depths. The ranges of ±standard deviation of measured data are shown by the error bars.

This negative association arises due to the fact that the coupling strength through coupling channels $p_4$ or $q_4$ hinge on $h$. Recalling the tight-binding model, we know that the supermodes dispersion shows strong dependence on the coupling strength between adjacent Tamm interface states. With the increase of etching depth $h$, the Tamm interface state will become more confined and, hence, the coupling strength will be reduced. As a result, the bandwidth of supermodes is decreased and the supermodes dispersion becomes more flat, indicating a high local density of optical states that may bring new inspiration about topological Tamm plasmon lasers[11-13, 33-35] or other applications requiring strong field confinement.[61-63]

**Conclusion**

In summary, we have proposed and demonstrated experimentally a TTS based on TPCs belonging to different class of topology. The bandgap uniquely induced by the folding of TPC's dispersion is unambiguously verified by angle-resolved reflection spectra. Through the coupling of Tamm interface states existing between TPCs, the TTS supporting hybrid supermodes is acquired. Moreover, we have attained high tunability of the bandwidth of supermodes via a varied etching depth. Compared with traditional Tamm states that confine at the interface between PC and a metal slab but homogeneous in the interface plane, the supermodes are hybrid collective modes and hold the uniqueness as nonhomogeneous and tunable localized Tamm states in the interface plane. These kind of supermodes may bring new insight into many applications, including Tamm plasmon laser[11-13], optical sensing,[14,15] selective thermal emitters,[16-18] and so on. Besides the etching depth, we can also change the number of TPCs between adjacent Tamm interface states to tune the dispersion of supermodes.[59] The flexible tailoring on the geometric parameters of double-layer metasurfaces and multilayered photonic structures indicates that the TTS is a versatile platform for the manipulation of Tamm plasmons. In addition, the TPC and the TTS can also be extended to two dimensional systems, in which new topological effects such as high-order topology[64] can be explored. Our work opens a new paradigm for research on topological phenomena of TPPs, which is instructive for the design of Tamm-based nanophotonic devices and has prospective applications for enhancing light-matter interactions.

**Methods**

*Synthesis of Tamm photonic crystals*. The photonic crystal, including $HfO_2/SiO_2$ multilayers is deposited on the $SiO_2$ substrate by Electron Beam Evaporation (AdNaNotek EBS-150U). After depositing a 25 nm-thick $Si_3N_4$ film on the photonic crystal using standard plasma-enhanced chemical-vapor deposition (Oxford PE100), we patterned a 30 nm thick silver film on the top of the $Si_3N_4$ by magnetron sputtering (GATAN 682PECS). Then the grating is etched from the silver film by a focused ion beam (FIB dual-beam FEI Helios 600i, 30 keV, 40 pA). The period length is $L = 400$ nm and the etching depth is $h$ which can be set different values to by tuning the etching parameters. Finally, we deposit a 25 nm thick silver film on gratings. The total area of the sample is 40 um×40 um.

*Characterizations of the Samples.* The setup for the angle-resolved microspectroscopy system is shown in Figure 1c (ARMS, Ideaoptics Inc.). The light source is a xenon lamp coupled into a scanning fiber. The incident light is focused by the objective lens (100×) with a numerical aperture of 0.9, and the region of measurement is 40 um×40 um limited by the diaphragm. The reflected light is collected and coupled into another scanning fiber which connects a spectrometer. Through the programmed angle-resolved mode of ARMS, we can get a series of reflection spectra under different angles.

*Dispersion of Tamm Photonic Crystals*. To calculate band structures of Tamm photonic crystals for TE polarizations, numerical simulations are performed by the eigenfrequency solver in COMSOL Multiphysics. The angle-resolved spectrum simulated results are obtained via the 3D Finite Difference Time Domain method (FDTD, Lumerical Inc.). The Broadband Fixed Angle Source Technique (BFAST) method in FDTD calculation is utilized to simulate oblique illumination with a broadband source. The varied incident angle from 0º to 60º and a broadband source from 380 to 600 THz are introduced to get the far-field intensity. For all the simulations, the polarization direction of the source is parallel to the y-axis in Figure 1.

**Supporting Information**

Band structures and associated Zak phases of two TPCs ($p$ and $q$); band structures of TPCs and TTSs with etching depth $h$ varying from 60nm to 100nm; band structures of TTS with the unit dimer composed of TPCs $p_2$-$q_6$, TPCs $p_4$-$q_4$ and TPCs $p_6$-$q_2$; the reduction of radiation losses by double-layer metal gratings.

**Author Contributions**

T.Q. and X.J. performed the sample fabrication and measurement, T.Q. performed numerical simulations. H.L. Q.W. and M.H. provided helpful discussions. T.Q. M.H and H.L. wrote the manuscript. H.L. and S.N.Z. initiated the program and directed the research. The manuscript was written through contributions of all authors. All authors have given approval to the final version of the manuscript.

The authors declare no competing financial interest.

**Acknowledgments**

The authors acknowledge Meng Xiao for helpful discussions and Micro/nano optical characterization and measurement center of NJU on the experimental measurements. This work was supported by the National Key Projects for Basic Researches of China (Grants No. 2017YFA0205700 and No. 2017YFA0303700), and the National Natural Science Foundation of China (Grants No. 11690033, No. 61425018, No. 11621091, and No. 11374151).

(26) Bruckner, R.; Zakhidov, A. A.; Scholz, R.; Sudzius, M.; Hintschich, S. I.; Frob, H.; Lyssenko, V. G.; Leo, K. Phase-locked coherent modes in a patterned metal-organic microcavity. *Nat. Photonics* **2012**, 6, 322.

(27) Dyer, G. C.; R. Aizin, G.; Allen, S. J.; Grine, A. D.; Bethke, D.; Reno, J. L.; Shaner, E. A. Induced transparency by coupling of Tamm and defect states in tunable terahertz plasmonic crystals. *Nat. Photonics* **2013**, 7, 9205.

(28) Ferrier, L.; Nguyen, H. S.; Jamois, C.; Berguiga, L.; Symonds, C.; Bellessa, J.; Benyattou, T. Tamm plasmon photonic crystals: From bandgap engineering to defect cavity. *APL Photonics* **2019**, 4, 106101.

(29) Lu, L.; Joannopoulos, J. D.; Soljačić, M. Topological photonics. *Nat. Photonics* **2014**, 8, 821.

(30) Khanikaev, A. B.; Shvets, G. Two-dimensional topological photonics. *Nat. Photonics* **2017**, 11, 763–773.

(31) Ozawa, T.; Price, H. M.; Amo, A.; Goldman, N.; Hafezi, M.; Lu, L.; Rechtsman, M. C.; Schuster, D.; Simon, J.; Zilberberg, O.; Carusotto, I. Topological photonics. *Rev. Mod. Phys*. **2019**, 91, 015006.

(32) Wang, Z.; Chong, Y.; Joannopoulos, J.; Soljačić, M. Observation of unidirectional backscattering-immune topological electromagnetic states. *Nature* **2009**, 461, 772.

(33) St-Jean; P.; Goblot, V.; Galopin, E.; Lemaître, A.; Ozawa, T.; Le Gratiet, L.; Sagnes, I.; Bloch, J.; Amo, A. Lasing in topological edge states of a one-dimensional lattice. *Nat. Photonics* **2017**, 11, 651.

(34) Bandres, M. A.; Wittek, S.; Harari, G.; Parto, M.; Ren, J.; Segev, M.; Christodoulides, D. N.; Khajavikhan, M. Topological insulator laser: Experiments. *Science* **2018**, 359, 6381.